\documentclass[preprint,superscriptaddress]{revtex4-1}

\usepackage{epsfig}
\usepackage{graphics}
\usepackage{amsmath}
\usepackage{color}

\newcommand{\ket}[1]{\left\vert#1\right\rangle}
\newcommand{\bra}[1]{\left\langle#1\right\vert}

\newcommand{\ppsi}{\left\vert \psi \right\rangle}

\begin{document}

\author{Yoav Sagi}
\author{Ido Almog}
\author{Nir Davidson}
\affiliation{Department of Physics of Complex Systems, Weizmann Institute of Science, Rehovot 76100, Israel }

\title{Process tomography of dynamical decoupling in a dense optically trapped atomic ensemble}

\begin{abstract}
Atomic ensembles have many potential applications in quantum information science. Owing to collective enhancement, working with ensembles at high densities increases the overall efficiency of quantum operations, but at the same time also increases the collision rate and leads to faster decoherence. Here we report on experiments with optically trapped $^{87}Rb$ atoms demonstrating a 20-fold increase of the coherence time when a dynamical decoupling sequence with more than 200 pi-pulses is applied. We perform quantum process tomography and demonstrate that using the decoupling scheme a dense ensemble with an optical depth of 230 can be used as an atomic memory with coherence times exceeding 3 sec.
\end{abstract}
\maketitle

Cold atomic ensembles can be used as an interface between matter and photonic qubits in quantum networks \cite{Duan2001,Lukin2003}, and in recent years vast experimental advances in this direction have been reported \cite{Kuzmich2003,Chou2005,Yuan2008,Zhao2009,Zhao_R_2009,schnorrberger:033003,zhang:233602}. Quantum information, which is mapped into the coherence between two atomic internal states, is gradually lost due to inhomogeneities and fluctuations in the energy difference between these states. For trapped atoms the inhomogeneities are caused by differential light shift in optical traps \cite{grimm2000} or by differential Zeeman shifts in magnetic traps, and by mean-field density dependent interaction shifts \cite{Harber2002}. Fluctuations arise due to collisions which are inherent to the high densities required to achieve a good overall efficiency of quantum operations \cite{PhysRevLett.82.4611,gorshkov2007}. Though fluctuations at low frequencies can be overcome by a single population inverting pulse - the celebrated coherence echo technique \cite{Hahn1950,Andersen2003}, as the collision rate increases this is no longer possible due to higher frequency components. Dynamical decoupling theories generalize this technique to multi-pulse sequences by harnessing symmetry properties of the coupling Hamiltonian \cite{Viola1998,PhysRevLett.82.2417,PhysRevLett.85.2272,uhrig2007,cywinski:174509}. Though dynamical decoupling was demonstrated in several experiments \cite{Fortunato2002,PhysRevLett.95.030506,Morton2006,PhysRevLett.103.040502,Biercuk2009,Du2009}, its use with atomic ensembles remains unexplored. In addition to its practical importance, this exploration is of theoretical interest since in trapped atomic ensembles the energy distribution is non-Gaussian and the fluctuations originate in self-interactions and not in a noisy external environment.

%


We consider atoms trapped in a conservative optical potential. The effective single particle Hamiltonian for atoms with internal states designated by $\ket{1}$ and $\ket{2}$ is given by
\begin{equation}\label{Hamiltonian}
\hat{H}=\hbar\left[\omega_0+\delta(t)\right]\ket{2}\bra{2}+\hbar\Omega(t)\ket{2}\bra{1}+h.c. \ \ ,
\end{equation}
where $\omega_0$ is the free space transition frequency between the states, $\delta(t)$ is a random frequency detuning sequence whose nature is determined by the potential inhomogeneities and collisions, and $\Omega(t)$ is the external control field which is used for the dynamical decoupling. Starting with an initial state $\ket{\psi(0)}=2^{-1/2}(\ket{1}+\ket{2})$ and no external control fields, the wave-function at any given time is given in the rotating frame by $\ket{\psi(t)}=2^{-1/2}(\ket{1}+e^{-i\phi(t)}\ket{2})$, where the phase difference is given by $\phi(t)=\int_0^\infty\delta(t)dt$. A schematic plot of three realizations of $\phi(t)$ is given in Fig. 1a (top), and it can be seen that the phase difference is accumulated in a constant rate between collisions \cite{kuhr:023406}. The ensemble coherence is characterized by the function $C(t)=\frac{|\langle \rho_{12}(t) \rangle|}{|\langle \rho_{12}(0) \rangle|}$, where $\rho_{12}$ is the off-diagonal element of the reduced two-level density matrix \cite{cywinski:174509}. As an example, for a Gaussian phase distribution, $P_\phi$, with a standard deviation $\sigma_\phi$ we obtain $C(t)=e^{-\frac{\sigma_{\phi}^2(t)}{2}}$, which shows that the coherence decays as the width of the phase distribution increases. The effect of a population inverting pulse ($\pi$-pulse) is to change the sign of $\delta$, and a train of such pulses lead to a much narrower phase distribution and slower decoherence, as depicted in Fig. 1a (bottom).

The experiments are carried out with cold $^{87}Rb$ atoms trapped in a far-off-resonance laser (see Fig. 1b, and more details in the supplementary information). The two relevant internal states are \mbox{$\ket{1}=\ket{F=1;m_f-1}$} and \mbox{$\ket{2}=\ket{F=2;m_f=1}$} in the $5^2S_{1/2}$ manifold, which are, to first order, Zeeman insensitive to magnetic fluctuations in the applied magnetic field of $3.2$G \cite{Harber2002}. The external control is done by means of two-photons transition (MW-RF photons), and the detection is state sensitive \cite{Khaykovich2000}. By gradually lowering the trapping laser intensity we reach the experimental conditions, at which we have $275,000$ atoms at a temperature of $1.7\mu K$, phase space density of $0.04$ and an average collision rate of $100s^{-1}$. The typical inhomogeneous decay time as measured in a Ramsey-like experiment is $\sim 150ms$. The peak optical depth for a non-polarized resonant light is $\sim230$.

\begin{figure}
\center\includegraphics[width=16cm]{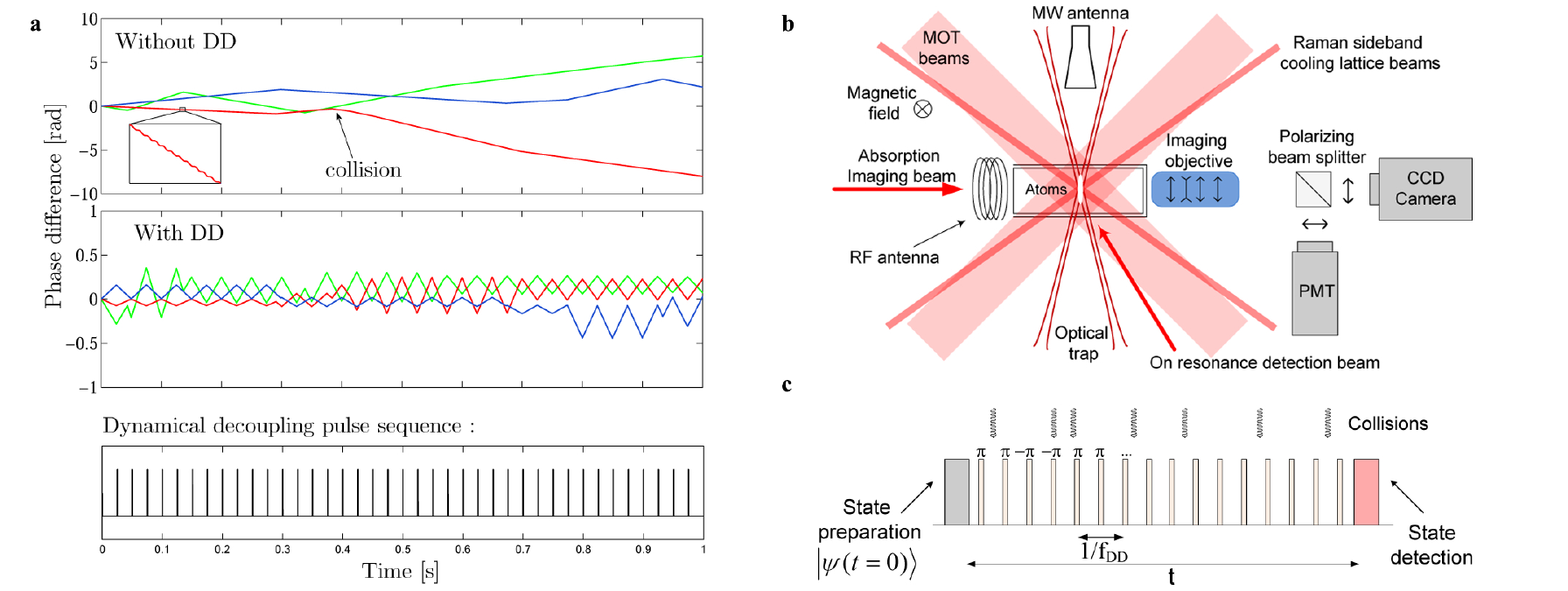}
    \caption{The experimental setup and dynamical decoupling sequence. \textbf{a}, Schematic drawing of the evolution of the relative phase between the two atomic internal states with and without the dynamical decoupling pulse sequence. Without dynamical decoupling (top), after a collision the average potential energy of the atom is changed and therefore also its average rate of phase accumulation, $\delta$. The inset shows the small oscillations due to the fast periodic atomic motion in the trap. Since the oscillation period is shorter than all relevant timescales in our experiment, we shall consider only $\delta(t)$ which is averaged over several oscillations \cite{kuhr:023406}. The dynamical decoupling pulse sequence is plotted in the lower graph, where each pulse is a $\pi$ pulse. With dynamical decoupling pulse sequence (plotted in the bottom), the spread of the phases is much smaller (note the different graph scales). \textbf{b}, The experimental apparatus. We laser cool $^{87}Rb$ atoms and trap them in a crossed red-detuned laser beams configuration. We employ state sensitive detection using a detection beam and a photomultiplier tube (PMT), and measure the density and temperature using absorption imaging on a CCD camera. \textbf{c}, The experimental pulse sequence starts with a state preparation, followed by a train of population inverting pulses with alternated phases $\{\pi,\pi,-\pi,-\pi,...\}$ to minimize the accumulation of errors due to pulse width and frequency inaccuracies, and a final state detection. The duration of each $\pi$ pulse is $\sim 0.5ms$ and its average fidelity is $\sim0.995$.}
\end{figure}

We employ a  Carr-–Purcell-–Meiboom-–Gill (CPMG) decoupling scheme \cite{RevModPhys.76.1037} and show in what follows that for collisional detuning fluctuations it is virtually optimal. The pulse sequence is composed of $n$ $\pi$-pulses at times $t_k=\frac{2k-1}{2n}t$ where $k=1\ldots n$ (see Fig. 1c), and we characterize it by the effective frequency $f_{DD}=\frac{n}{2t}$. We study the effect of the dynamical decoupling scheme by performing a quantum process tomography (QPT) \cite{QCQI}. QPT enables us to reconstruct the $\chi$-matrix which gives a convenient way to calculate the density matrix after the process, $\rho_{out}$, in terms of the initial density matrix, $\rho_{in}$, by $\rho_{out}=\mathcal{E}[\rho_{in}]=\sum_{k,l}\hat{E}_k\rho_{in}\hat{E}^\dagger_l \chi_{kl}$, where $\hat{\mathbf{E}}=(\hat{I},\hat{X},-i\hat{Y},\hat{Z})$ with $(\hat{I},\hat{X},\hat{Y},\hat{Z})$ being the Pauli matrices. In the experiment we start with a set of initial states after which we apply the decoupling scheme and measure $\rho_{out}$ by quantum state tomography (for more details see the supplementary information). The results of a QPT of a dynamical decoupling sequence with $f_{DD}=35$Hz is depicted in Fig. 2a. There are two distinctive decay timescales for the equatorial plane and the $z$-axis, which corresponds to phase damping noise processes and depolarizing noise processes ($T_1$), respectively. The former originates from the fluctuations in $\delta$ and it is the dominant noise process which determines the ensemble coherence time, $\tau_c$, and is quantified by $C(t)$. Depolarization process is induced by inelastic collisions \cite{Widera2006}, and its typical timescale is measured to be $T_1=6s$. The worst case fidelity of the ensemble as a quantum memory, defined as $\mathcal{F}=\underset{\ket{\psi}}{\min}{\langle \psi |\mathcal{E}\left[\ket{\psi}\bra{\psi}\right]|\psi\rangle}$, is calculated from the measured $\chi$-matrix to be $\mathcal{F}=0.83$, $0.74$ and $0.64$ for $1$, $2$ and $3$ seconds, respectively, which corresponds to an exponential decay timescale of $\tau_c=2.4sec$. The contraction of the Bloch sphere is symmetric in the equatorial plane, which indicates that the decoupling scheme is insensitive to the stored superposition. We demonstrate this point with a direct measurement in which we start with two orthogonal initial states in the equatorial plain:  $\ket{\psi_1}=\frac{1}{\sqrt{2}}(\ket{1}+\ket{2})$ and $\ket{\psi_2}=\frac{1}{\sqrt{2}}(\ket{1}+e^{i\pi/2}\ket{2})$. For both states we scan the phase of a final $\pi/2$ pulse added to the sequence and measure the population at $\ket{2}$ normalized to its initial value. The results depicted in Fig. 2b exhibit the same contrast and preserve the $\pi/2$ phase shift between the two initial states.

\begin{figure}
\center\includegraphics[width=16cm]{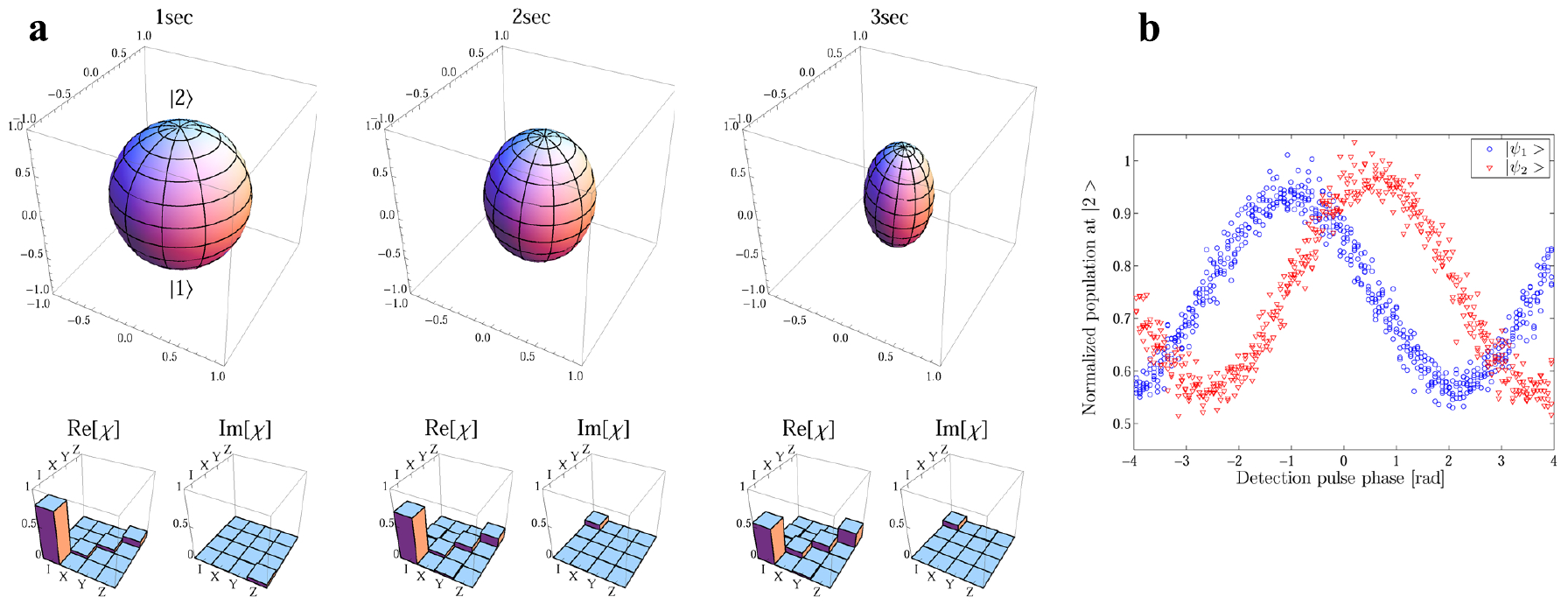}
    \caption{Quantum process tomography (QPT) of dynamical decoupling in cold atomic ensemble. \textbf{a}, We perform QPT of dynamical decoupling with $f_{DD}=35$Hz and reconstruct the $\chi$-matrix which defines the process $\mathcal{E}[\rho]$. Any single-qubit density matrix $\rho$ can be mapped to a point in space $\bar{r}(\rho)=(tr(\hat{X}\rho),tr(\hat{Y}\rho),tr(\hat{Z}\rho))$, with $(\hat{X},\hat{Y},\hat{Z})$ being the Pauli matrices. The colors and lines are chosen for the initial states, $\rho_{in}$, which lie on a sphere with a radius $1$. For each of these states we calculate the process outcome, $\rho_{out}=\mathcal{E}[\rho_{in}]$, and plot it with its initial color at $\bar{r}(\rho_{out})$. The contraction of the Bloch sphere is more pronounced on the equatorial plane which shows that the main noise process is phase damping. Slower depolarizing process ($T_1$) which uniformly shrinks the sphere are also apparent. This can be also seen in higher values of $\chi_{zz}$ compared to $\chi_{xx}$ and $\chi_{yy}$. The fidelity of the ensemble as a quantum memory is calculated to be $0.83$, $0.74$ and $0.64$ after $1$, $2$ and $3$ seconds, respectively, yielding a coherence time of $\tau_c=2.4$ sec. There is also a rotation of the sphere around the $\ket{1}$-$\ket{2}$ axis at a rate of $\sim9^\circ sec^{-1}$ due to small inaccuracies in the control field. \textbf{b}, Storage of two orthogonal initial states in the equatorial plane: $\ket{\psi_1}=\frac{1}{\sqrt{2}}(\ket{1}+\ket{2})$ and $\ket{\psi_2}=\frac{1}{\sqrt{2}}(\ket{1}+e^{i\pi/2}\ket{2})$. We apply the decoupling scheme and add to it another pulse independent of the initial state with a phase and duration that are chosen to correct for the small rotation of the Bloch sphere. We measure the population at $\ket{2}$ after $3$ sec, normalized to the initial population, versus the phase of a $\pi/2$ detection pulse. The phase difference between the two initial states is conserved and also the contrast which corresponds to a higher $\tau_c=3$ sec due to the correction of the rotation. The fringe contrast is not centered to $0.5$ due to inelastic $m$-changing transitions in the hyperfine level $F=2$.}
\end{figure}

The decay of the coherence with the dynamical decoupling pulse sequence and assuming a Gaussian phase distribution can be obtained in a system-reservoir framework \cite{cywinski:174509,gordon:010403}:
\begin{equation}\label{decay_equation}
C(t)=e^{-\int_0^\infty d\omega{S_\delta(\omega)F(\omega t)}/{\pi\omega^2}} \ \ ,
\end{equation}
where the argument is the overlap integral between the fluctuations power spectrum, $S_\delta(\omega)=\int^\infty_{-\infty} \langle \delta(t) \delta(0) \rangle e^{i\omega t}dt$, and a filter function which encapsulates the information on the dynamical decoupling pulse sequence and is given by $F(\omega t)=\frac{1}{2}|\sum_{k=0}^n(-1)^k (e^{i\omega t_{k+1}}-e^{i\omega t_{k}})|^2$ with $t_0=0$ and $t_{n+1}=t$. For each atom $\delta(t)$ is a sequence of constant detunings connected by ``jumps'' which occur after each collision. Since the collision times follow Poisson statistics, the detuning correlation function decay exponentially $\Phi_\delta(t)=\langle \delta(t+0) \delta(0) \rangle=\sigma_\delta^{2} e^{-\Gamma |t|}$, where $\Gamma^{-1}$ is the correlation time of the detuning and $\sigma_\delta$ is the standard deviation of the detunings distribution. The power spectrum is given by $S_\delta(\omega)=\int^\infty_{-\infty} \Phi_\delta(t)e^{i\omega t}dt=\frac{2\Gamma\sigma_\delta^2}{\Gamma^2+\omega^2}$. By solving numerically Eq. (\ref{decay_equation}) and leaving the $\{t_i\}_{i=1}^{n}$ as free parameters, we find that the optimal decoupling sequence for a Lorentzian power spectrum is given by $t_i=\frac{\eta+i-1}{n-1+2\eta}t$, where $i=1...n$ and $0.5\le \eta \le 1$ is a numerical factor which depends on $n$ and $t$. For $\frac{\Gamma t}{n}\ll 1$ we find $\eta\approx0.5$, for which we retrieve the CPMG pulse sequence. Furthermore, even when $\frac{\Gamma t}{n}\approx 1$ the coherence time with the CPMG pulse sequence differs by less than $1\%$ from the optimal value. We have tested theoretically and experimentally other dynamical decoupling schemes, and in particular the one suggested in Ref. \cite{uhrig2007}, and verified that they are indeed inferior to the CPMG sequence in our Lorentzian fluctuations power spectrum (for more details see the supplementary information).

We measure $C(t)$ directly by the following sequence: a short $\pi/2$ pulse with a phase $\varphi$ prepares the atoms in the superposition $\ppsi=\frac{1}{\sqrt{2}}(\ket{1}+e^{i\varphi}\ket{2})$, following are the decoupling pulses and finally the coherence is measured from the length of the Bloch vector measured with a quantum state tomography. Though $C(t)$ does not have to follow, a priori, some well defined function, the experimental results depicted in Fig. 3a show that the data is well fitted by an exponentially decaying function $e^{-t/\tau_c}$. The exponential decay is expected in the Markovian limit, where the decay timescale is much larger than the fluctuations correlation time \cite{cywinski:174509}. This is indeed the case in our experiment, as can be seem in the inset of Fig. 3a where we calculate $C(t)$ using Eq. (\ref{decay_equation}) with several values of $f_{DD}$; The lines on a semilog scale show deviations from linear fits only for $t\ll \Gamma^{-1}$. We also measure the dependance of the coherence time on the dynamical decoupling pulse rate and the results are shown in Fig. 3b. We observe a quadratic increase of the coherence time versus $f_{DD}$ up to $35$Hz, where we obtain a $20$-fold improvement with coherence times exceeding 3 sec.

\begin{figure}
\center\includegraphics[width=16cm]{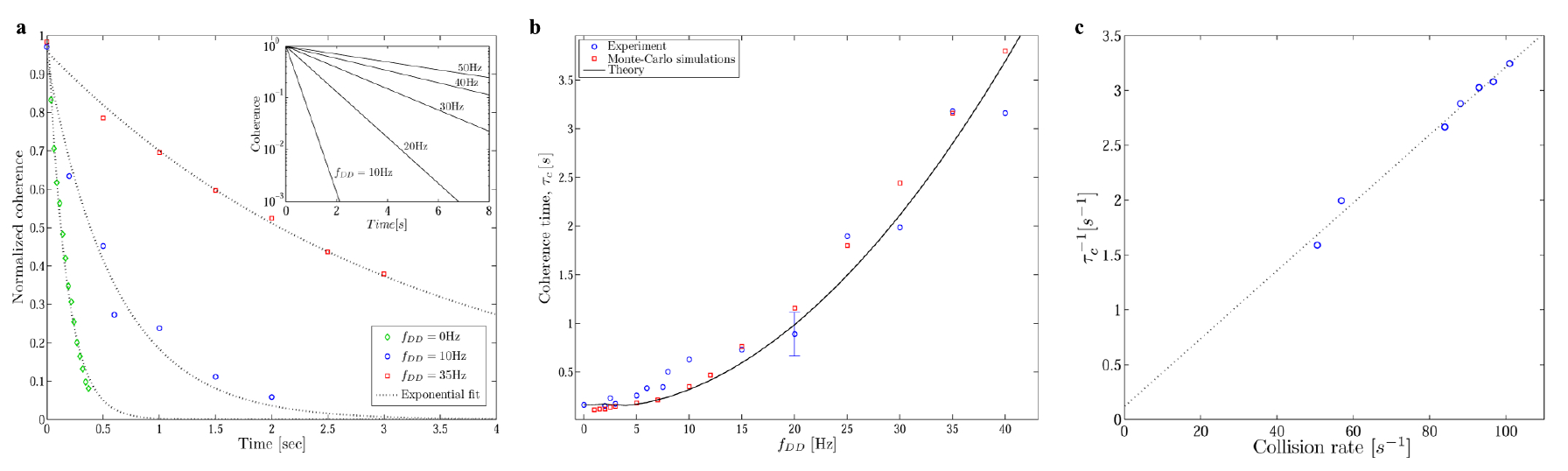}
   \caption{The ensemble coherence time with a CPMG dynamical decoupling pulse sequence. \textbf{a}, The coherence left in the ensemble without dynamical decoupling ($f_{DD}=0$Hz) and with two representing dynamical decoupling pulse rates after different waiting times, normalized to the initial stored coherence. From the fit to the exponential decay function (dotted line) we extract the coherence time. The inset shows calculations according to Eq.(\ref{decay_equation}) of the coherence time in a semilog scale, and demonstrates the expected exponential decay behavior. \textbf{b}, The parabolic dependence of the coherence time on the dynamical decoupling pulse rate. The experimental data (blue circles) agrees well both with the theoretical prediction of Eq.(\ref{decay_equation}) taking a Lorentzian power spectrum (solid line) with $\sigma_\delta=23.8sec^{-1}$ and $\Gamma=37.5sec^{-1}$ which were extracted from measured quantities and with no fit parameters, and with molecular dynamics Monte-Carlo simulations done with $1000$ atoms in a 3D harmonic trap with an inhomogeneous decay time similar to the experiment (red squares). The errorbar is a estimated from the fits shown in (a). \textbf{c}, The dephasing rate is measured for different collision rates, demonstrating the linear dependence of Eq.(\ref{coherence_time}). These experiments were done with $f_{DD}=8$Hz.}
\end{figure}

In order to explain these results we present a qualitative model for the coherence time. Without collisions the inhomogeneous dephasing rate is proportional to $\sigma_\delta$. For simplicity we assume that if a collision did not occur between two consecutive $\pi$-pulses the inhomogeneous broadening is averaged out. If a collision occurred, however, the width of the ensemble phase distribution increases by $\sim f_{DD}^{-1} \sigma_\delta$. The number of collisions up to a time $t$ is $\Gamma_{col} t$, and since we add random variables (i.e. the accumulated phase), the width of the phase distribution increases as a square root of time: $\Delta \Phi(t) \sim f_{DD}^{-1} \sigma_\delta \sqrt{\Gamma_{col} t}$. For cold collisions in 3D harmonic trap, the relation between the collision rate and the relaxation rate was shown to be\cite{Monroe1993} $\Gamma_{col}=2.7\cdot \Gamma$. The coherence time, $\tau_c$, is the time for which the width of phase distribution is on the order of $1$, yielding
\begin{equation}\label{coherence_time}
\tau_c \sim f_{DD}^{2} \sigma_{\delta}^{-2} \Gamma^{-1} \ \ ,
\end{equation}
with a parabolic dependence on $f_{DD}$. This result can be also obtained from Eq.(\ref{decay_equation}) by approximating $ \frac{F(\omega t)}{\pi (\omega t)^2}\approx\delta_{Dirac}(\omega t-2\pi f_{DD} t)$ and using the Lorentzian power spectrum.

Exact calculations of $\tau_c$ using Eq.(\ref{decay_equation}) without fitting parameters are presented in Fig. 3b in good agreement with the experimental data. The calculations are done with a Lorentzian power spectrum where the parameter $\Gamma$ and $\sigma_\delta$ are calculated from measured quantities. $\Gamma$ is extracted from the collision rate which is calculated using the measured temperature, number of atoms and trap oscillation frequencies. The parameter $\sigma_\delta$ is measured in a Ramsey experiment at very low densities, where the collisions can be disregarded and $\sigma_\delta$ can be extracted from the measured dephasing rate (for more details see the supplementary information). We also perform Monte-Carlo simulations, where we solve for the classical motion of atoms in the trap including collisions, and calculate the Ramsey signal by tracing each atom's accumulated phase along its trajectory. The results of the simulations are also depicted in Fig. 3b, and agree well with both theory and experiments. We conclude that the effect of collisions can be indeed formulated as an effective single spin Hamiltonian coupled to a reservoir. Moreover, although the detunings of atoms trapped in a $3D$ harmonic trap are not normally distributed (for more details see the supplementary material), the distribution of their accumulated phase can be well approximated by a Gaussian owing to the validity of the central limit theorem and large number of collisions involved.



Another outcome of Eq.(\ref{coherence_time}) is the linear dependence of the coherence time on $\Gamma^{-1}$. In the experiment we change $\Gamma$ by reducing the density and collision rate while keeping the temperature, and therefore $\sigma_\delta$, unchanged \cite{kuhr:023406}. This is accomplished by reducing the intensity of the cooling lasers in the trap loading phase. In Fig. 3c we plot $\tau_c^{-1}$ versus the average collision rate for a pulse rate of $f_{DD}=8$Hz. As expected, the coherence time is inversely proportional to the collision rate.

In conclusion, we have demonstrated that dynamical decoupling can substantially increase the coherence time of a dense optically trapped atomic ensemble. In the current work the ensemble was treated as an effective single spin system which accounts for storage schemes based on collinear EIT \cite{Lukin2003}. A natural extension is the application of dynamical decoupling to Raman scattering schemes in which a global coherence is created between the atoms \cite{Duan2001,Kuzmich2003}. Another promising prospect lies in novel hybrid approaches to quantum computation combining atomic ensembles and superconducting devices\cite{petrosyan:040304}, where the application of dynamical decoupling could reduce the error probability during the characteristic $100\mu s$ single qubit gate to less than $10^{-4}$ - below the current estimated threshold for a fault-tolerant quantum computation.

We thank Roee Ozeri, Guy Bensky, Goren Gordon, Gershon Kurizki and Rami Pugatch for helpful discussions. We acknowledge the financial support of MIDAS, MINERVA, ISF, and DIP.

%

\end{document}


\author{Yoav Sagi}
\author{Ido Almog}
\author{Nir Davidson}
\affiliation{Department of Physics of Complex Systems, Weizmann Institute of Science, Rehovot 76100, Israel }

\title{Supplementary information for the manuscript ``Process tomography of dynamical decoupling in a dense optically trapped atomic ensemble''}

\maketitle

\section*{The experimental setup}
Initially $\sim10^9$ atoms are trapped and cooled in a magneto-optical trap, and further cooled by Sisyphus and Raman sideband techniques. We use rapid adiabatic passage with constant RF radiation and a ramped magnetic field to transfer the atoms from state $\ket{5^2S_{1/2},F=1;m_f=1}$ ending with $80\%$of the atoms at $\ket{1}$, and the rest in state \mbox{$\ket{5^2S_{1/2},F=0;m_f=0}$}. The atoms are loaded into an optical dipole trap created by two horizontal crossing beams at an angle of $28^\circ$, creating an oval trap with an aspect ratio of 1:3.9 with a measured radial oscillation frequency of $2\pi \cdot 330$Hz. The $50\mu m$ waist laser beams originate from a single frequency Ytterbium fiber laser at $1064$nm. Their polarization is parallel to the magnetic field, and frequency differ by $120$MHz to eliminate standing waves. The external control is composed of RF radiation at $2.15$MHz and microwave radiation at $\sim6.8$GHz, both locked to an atomic standard. Another effect which should be taken into account is the levels shift induced by the MW field \cite{PhysRevLett.24.861}. We have carefully measured this shift, which in the maximum MW power reaches to $\sim50$Hz, and we take it into account when setting the frequency of the external fields. We carry out evaporative cooling for $2.5sec$ to a laser power of $0.16$W and back to a final value of $0.64$W. The spontaneous scattering rate is less than $1s^{-1}$ and the trap lifetime is longer than $5s$. We measure the temperature by an absorption imaging after a time of flight.

\section*{Quantum state tomography}
In quantum state tomography the super-operator matrix $\chi$ is determined by implementing a complete set of qubit input states. For a single qubit this is conveniently done by a measurement of $tr[\hat{X}\rho]$, $tr[\hat{Y}\rho]$ and $tr[\hat{Z}\rho]$ \cite{QCQI}. We measure the population at state $\ket{2}$ with and without another $\pi$ pulse, and the difference of the two values normalized to the initial population at state $\ket{1}$ gives $\frac{tr[\hat{Z}\rho]+1}{2}$. We obtain the other two projections by apply a $\pi/2$ pulse in the suitable axis and then measure $tr[\hat{Z}\rho]$.

\section*{Dephasing in a 3D harmonic trap}
The dephasing in a $3D$ harmonic trap without collisions is given by $C(t)=\left[1+\frac{\left(\sigma_\delta t\right)^2}{3}\right]^{-3/2}$ \cite{kuhr:023406}, and solving $C(\tau_1)=e^{-1}$ we obtain $\tau_1=\sqrt{3 \left(e^{2/3}-1\right)}\sigma_\delta^{-1}\approx 1.69 \cdot \sigma_\delta^{-1}$.

\section*{Optimal decoupling pulse sequence for a Lorentzian spectral function}
\begin{figure}
    \begin{center}
    \centerline{\includegraphics[width=10cm]{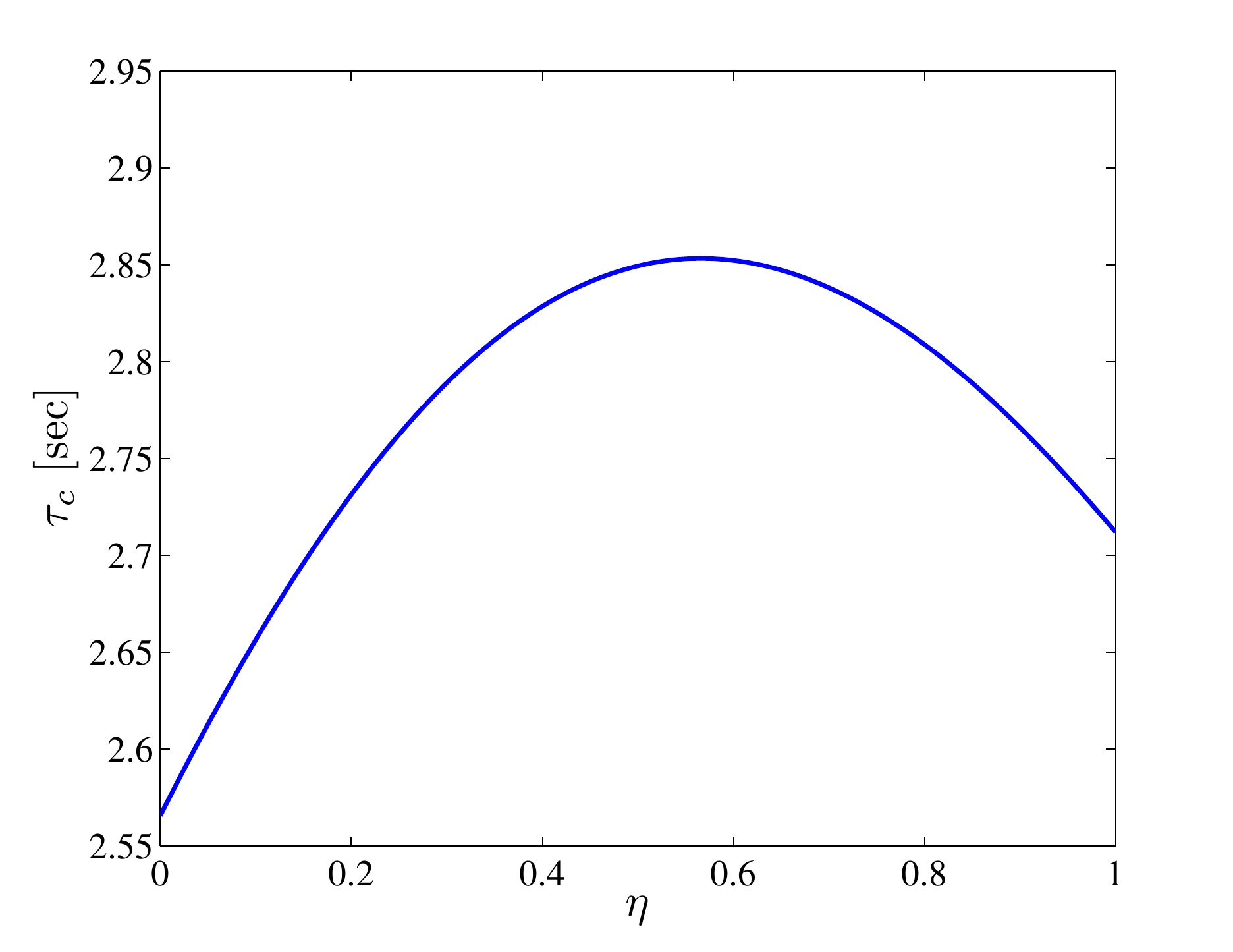}}
    \end{center}\caption{Coherence time versus $\eta$ for $f_{DD}=35$Hz and a lorentzian power spectrum with the typical experimental parameters $\Gamma=37.5s^{-1}$ and $\sigma_\delta=23.8s^{-1}$. The calculation of $\tau_c$ is done at $t=1s$.}\label{coherence_time_for_fdd_35hz}
\end{figure}

\begin{figure}
    \begin{center}
    \centerline{\includegraphics[width=10cm]{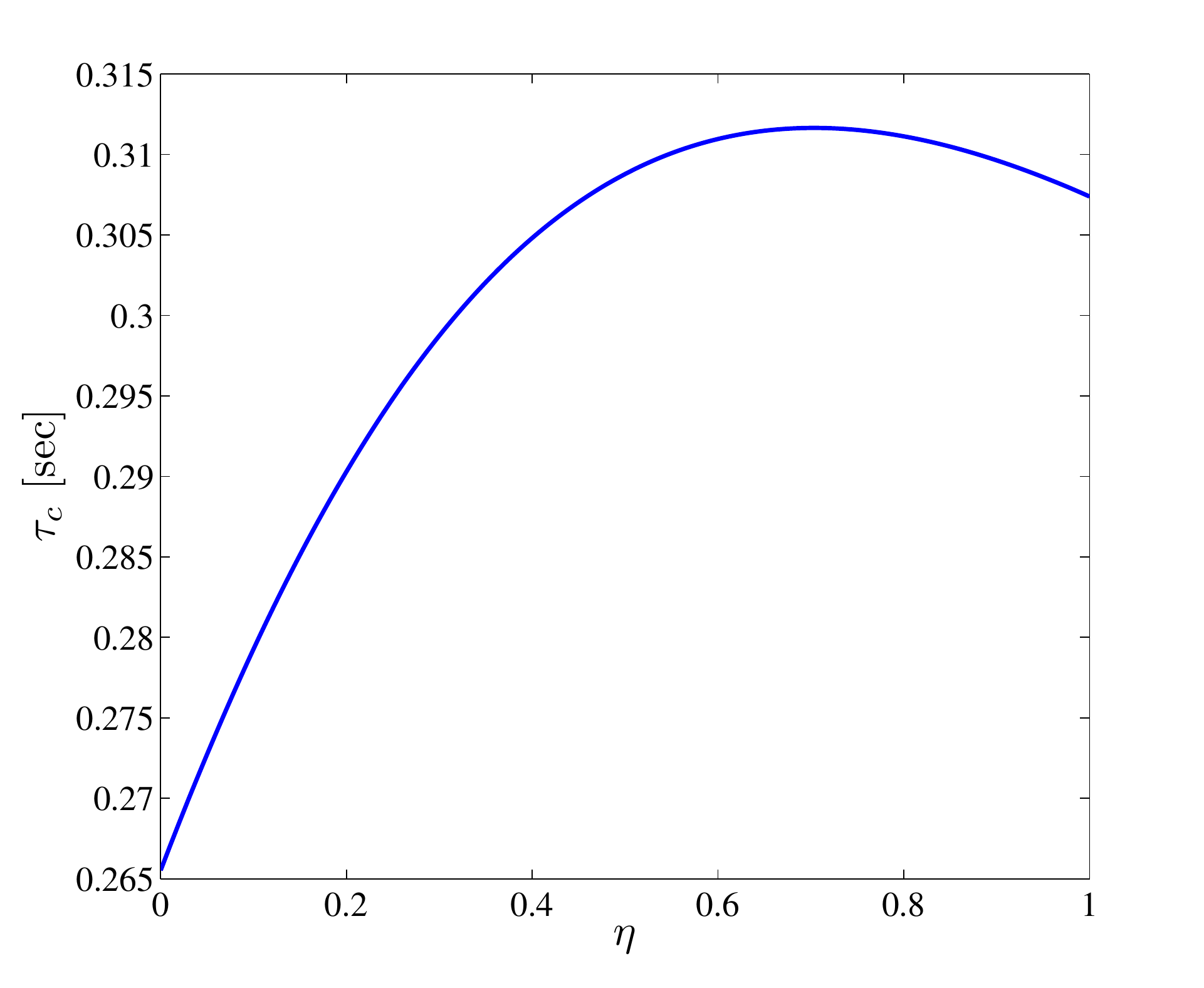}}
    \end{center}\caption{Coherence time versus $\eta$ for $f_{DD}=10$Hz and a lorentzian power spectrum with the typical experimental parameters $\Gamma=37.5s^{-1}$ and $\sigma_\delta=23.8s^{-1}$. The calculation of $\tau_c$ is done at $t=1s$.}\label{coherence_time_for_fdd_10hz}
\end{figure}

As explained in the text, we want to find the pulse sequence timing series $\{t_i\}_{i=1}^{n}$ which maximizes the coherence as calculated by Eq.(2). We use a nonlinear minimization method (Nelder-Mead) implemented in MATLAB software for the minimum decoherence. We have consistently obtained a time series close to the form given in the text: $t_i=\frac{\eta+i-1}{n-1+2\eta}t$, where $i=1...n$ and $0.5\le \eta \le 1$ is a numerical factor which depends on the parameters of the problem. For $f_{DD}>\Gamma$, the difference of the time series for the optimal $\eta$ to the one with $\eta=0.5$ (which corresponds to the CPMG scheme) is less then $1\%$. As an example we plot the coherence time as a function of $\eta$ for $f_{DD}=35$Hz in Fig. \ref{coherence_time_for_fdd_35hz}. The difference in this case between $\tau_c$ of $\eta=0.5$ and the optimal value is $\sim0.14\%$. In Fig. \ref{coherence_time_for_fdd_10hz} we plot the same calculation for $f_{DD}=10$Hz. Here the difference between the optimal value and the value at $\eta=0.5$ is $\sim0.93\%$. Therefore the value $\eta=0.5$ was chosen to be used in the experiment.

\begin{figure}
    \begin{center}
    \centerline{\includegraphics[width=10cm]{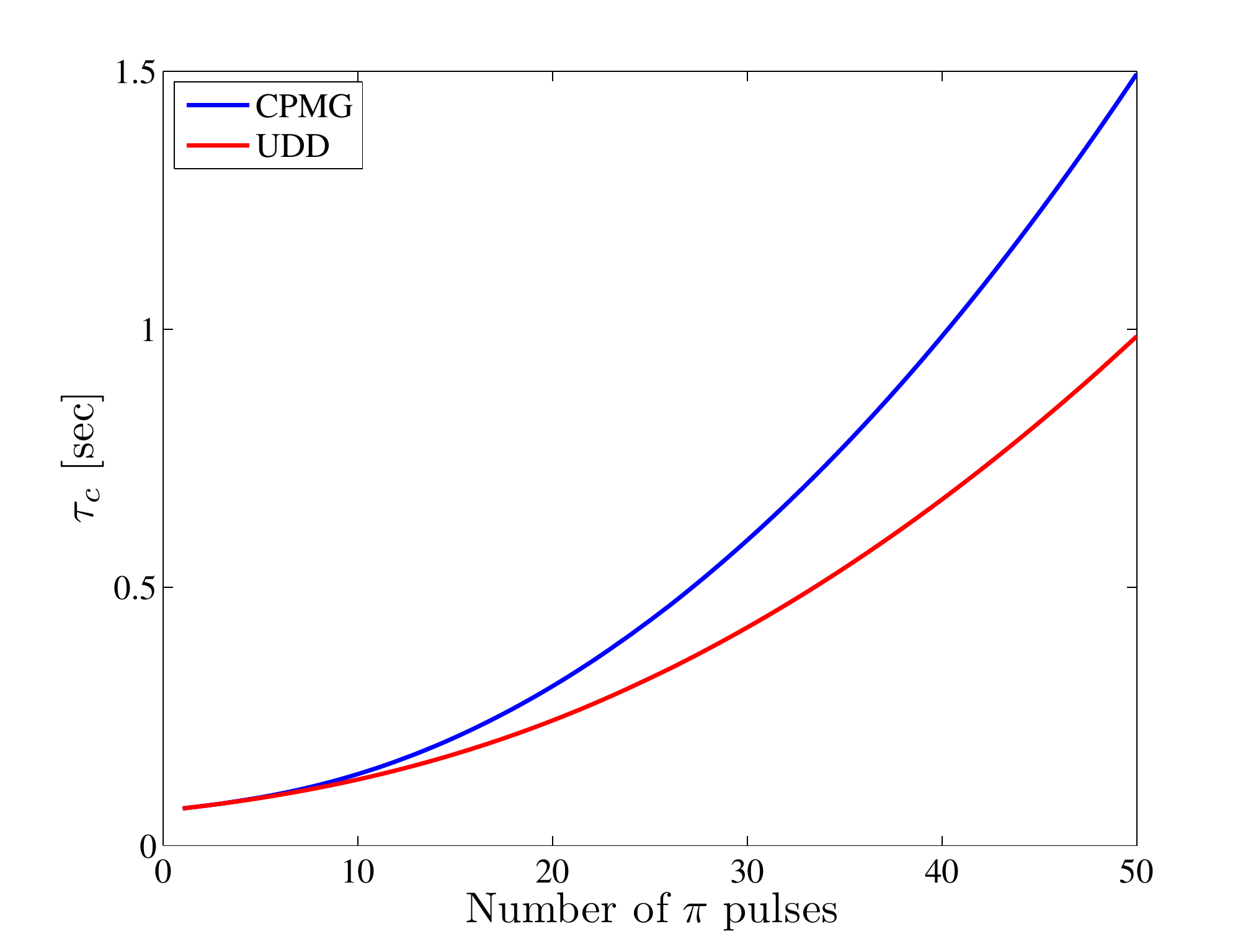}}
    \end{center}\caption{Coherence time versus the number of $\pi$=pulses for a CPMG dynamical decoupling scheme and the UDD scheme, for a lorentzian power spectrum with the typical experimental parameters $\Gamma=37.5s^{-1}$ and $\sigma_\delta=23.8s^{-1}$. The calculation of $\tau_c$ is done at $t=1s$.}\label{coherence_time_versus_number_of_pulses_UDD_and_CPMG}
\end{figure}

In regard to the Uhrig dynamical decoupling scheme (UDD) \cite{uhrig2007}, we plot in Fig. \ref{coherence_time_versus_number_of_pulses_UDD_and_CPMG} a comparison between CPMG and UDD as a function of the number of $\pi$-pulses. As can be clearly seen from the graph, the UDD scheme is inferior to the simpler CPMG scheme for a Lorentzian bath spectrum. This is not surprising since the UDD is expected to be superior in spectral functions that exhibit a cutoff in frequency. This point was already discussed in Ref. \cite{cywinski:174509}. We have also tested this experimentally for several number of pulses and consistently obtained coherence times with the UDD scheme which are $10\%$-$20\%$ lower than the coherence times obtained with the CPMG scheme.


%